# Design of an impedance matching acoustic bend

Yuzhen Yang[1], Han Jia[1,2], Wenjia Lu[1], Zhaoyong Sun[1] and Jun Yang[1,2]

[1] Key Laboratory of Noise and Vibration Research, Institute of Acoustics, Chinese Academy of Sciences, Beijing 100190, People's Republic of China

[2] State Key Laboratory of Acoustics, Institute of Acoustics, Chinese Academy of Sciences, Beijing 100190, People's Republic of China

E-mail: hjia@mail.ioa.ac.cn;   jyang@mail.ioa.ac.cn

## Abstract

We propose the design of an impedance matching acoustic bend in this article. The bending structure is composed of sub-wavelength unit cells with perforated plates and side pipes, whose mass density and bulk modulus can be tuned simultaneously. So the refraction index and the impedance of the acoustic bend can be modulated simultaneously to guarantee both the bending effect and the high transmission. The simulation results of sound pressure field distribution show that the bending effect of the impedance matching acoustic bend is very good. Transmission spectra of the impedance matching acoustic bend and the acoustic bend composed of perforated plates only are both calculated for comparison. The results indicate that the impedance matching acoustic bend is successful in improving the impedance ratio and increasing the transmission obviously.

# Introduction

Acoustic metamaterial basically covers all artificial materials with periodic or pseudoperiodic sub-wavelength unit cells, which get its response mainly from its geometrical structure, rather than its chemical components. By adjusting the structure parameters of acoustic metamaterials, we can obtain some extraordinary acoustic parameters which cannot realize easily with natural materials, e.g. negative mass density and bulk modulus,[1-5] anisotropic mass density tensor,[6,7] and anisotropic elasticity tensor.[8] In previous studies on acoustic metamaterials, various amazing phenomena have been discovered,[9-19] e.g. negative refraction,[9-11] rainbow trapping,[12-14] and acoustic cloaking.[15-19] And the acoustic bending effect is one of the interesting phenomena.

For conventional waveguide with a bend, the reflection from the side wall will distort the wave front and the wave modes in the waveguide will become chaotic. A lot of research works about optical bend[20-25] have been done to solve the problem. The research of acoustic bend is also important in various fields such as sound absorption in pipeline and detections in industry. L. Wu *et al.* employed a two-dimensional graded sonic crystal to realize an acoustic bending waveguide in a wide frequency range.[26] Y. Wang *et al.* designed an acoustic bending waveguide using anisotropic density-near-zero metamaterial.[27] The surface waves on the input and output interfaces of the anisotropic density-near-zero metamaterial induce the sound energy flow to be redistributed and match perfectly with the propagating modes inside the waveguide. M. Ghasemi Baboly *et al.* introduced an isolate, single-mode, 90º bend phononic crystal waveguide.[28] A phononic crystal consisting of an array of circular air holes in an aluminum substrate is used, and waveguides are created by introducing a line defect in the lattice.

In our previous research, an acoustic bend composed of perforated plates (ABPP) is proposed to manipulate the sound waves inside a two-dimensional waveguide.[29] However, there is reflection from the incident surface and exit surface of the ABPP for impedance mismatch. In this article, a design of an impedance matching acoustic bend (IMAB) to avoid the reflection from the side wall and enhance the energy transmission is proposed. We introduce new complex unit cells with perforated plates and side pipes whose mass density and bulk modulus can be tuned simultaneously. The structure satisfies both the refractive index distribution and the acoustic impedance successfully. As a result, the IMAB manipulates the sound propagation inside the waveguide effectively and keeps a high transmission in both a wide frequency range and a wide angle range.

## Theory and design

The proposed IMAB is an arc-shaped waveguide with an inner radius $r_1$, an outer radius $r_2$ and a bending angle $0.5\pi$ as shown in Fig. 1(a). The sound waves are supposed to travel in the circumferential direction (indicating by the bold solid curve with an arrow). It is known that the acoustic wave propagation can be manipulated by designing proper refractive index distribution of the media. In acoustic bending structure, the variation of phase keeps the same at different radial positions. And the refractive index distribution should be $n = A/r$.[21] Besides the refractive index, the acoustic impedance also influences the characteristics of sound propagation. Taking both refractive index and impedance match into consideration, the equations should be expressed as follows:

$$\begin{cases} n = \dfrac{A}{r} \\ Z = Z_0 \end{cases}, \qquad (1)$$

where $A$ is a constant and $r$ is the radial position of the acoustic bend; $Z_0$ is the acoustic impedance of air. From Eq. (1), it is observed that the refractive index changes along the radial direction while the impedance keeps constant. The bending effect of the acoustic bend is derived from the continuous variation of refractive index, which is inversely proportional to the radial position of the structure. The refractive index $n$ and the acoustic impedance $Z$ are functions of the relative mass density $\rho$ and the relative bulk modulus $B$ as shown in Eq. (2):

$$\begin{cases} n = \sqrt{\dfrac{\rho}{B}} \\ Z = \sqrt{\rho B} * Z_0 \end{cases} \quad (\rho = \rho'/\rho_0 \quad B = B'/B_0), \quad (2)$$

$\rho_0$, $B_0$ are the mass density and bulk modulus of air; $\rho'$, $B'$ are the mass density and bulk modulus of the IMAB. Combining Eqs. (1) and (2), the spatial distribution of relative mass density and relative bulk modulus can be expressed as follows:

$$\begin{cases} \rho = \dfrac{1}{r}, \\ B = r \end{cases} \quad (3)$$

here we choose $A = 1$ for simplicity. From Eq. (3), distributions of the mass density and the bulk modulus are functions of the radial position. The mass density is inversely proportional to the radius while the bulk modulus is directly proportional to the radius. Figure 1(b) shows the relationship curves for mass density and bulk modulus with the radius $r$. The filled triangle line stands for the variation of the mass density and the solid circle line stands for the variation of the bulk modulus.

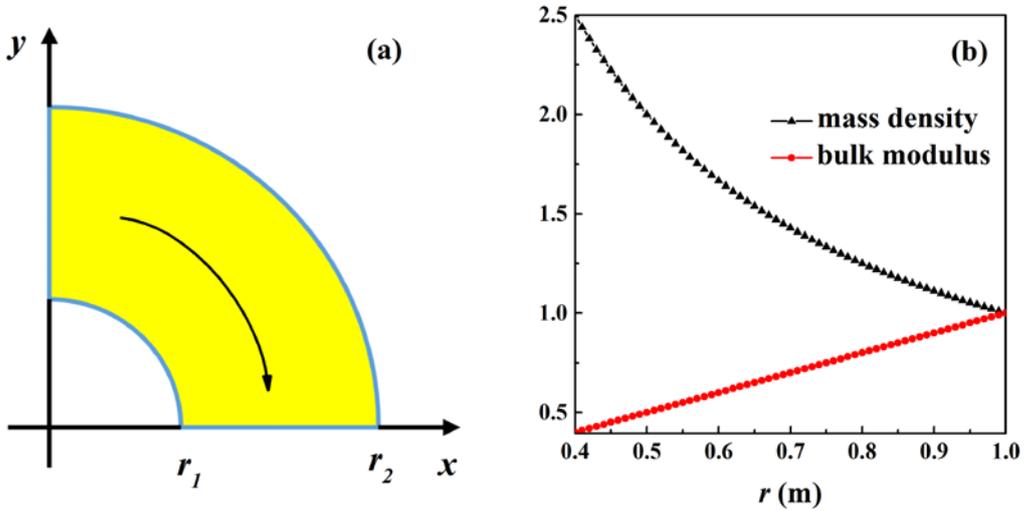

FIG. 1. (a) Schematic of the acoustic bend. The bold solid curve with arrow denotes the direction of wave propagation. (b) The relationship curves for mass density and bulk modulus with the radius r. The filled triangle line stands for the variation of the mass density and the sold circle line stands for the variation of the bulk modulus.

In our design, the inner and outer radii for the bending structure are $r_1 = 0.4$m and $r_2 = 0.99$m respectively, with the bending angle of $0.5\pi$. To realize the prototype by acoustic metamaterial, the bending structure is divided into 59 arc-shaped layers with the same width of 10mm along the radial direction because the mass density and the bulk modulus change with the variation of the radial position. Each layer is considered as a homogeneous material with the physical parameters equal to the value of the center line of the layer. According to the radial position of the center line, the mass density and the bulk modulus can be calculated by Eq. (3). Then each layer is divided into 30 periodical units along the circumferential direction. As a result, the whole structure is divided to $59 \times 30$ arc-shaped unit cells. The angle and the width of every unit cell is $3º$ and 10mm, respectively. The length $d$ of unit cell is average arc length, which is the function of radial position $r$. In fact, the practical structure of the acoustic bend is a quasi-two-dimensional structure with the height $h$ equal to 10mm, because these component unit cells have a certain height of 10mm. The arc-shaped unit cell could be considered as a cuboid when calculating acoustic parameters of the cell in the following because of the small angle, width and length.

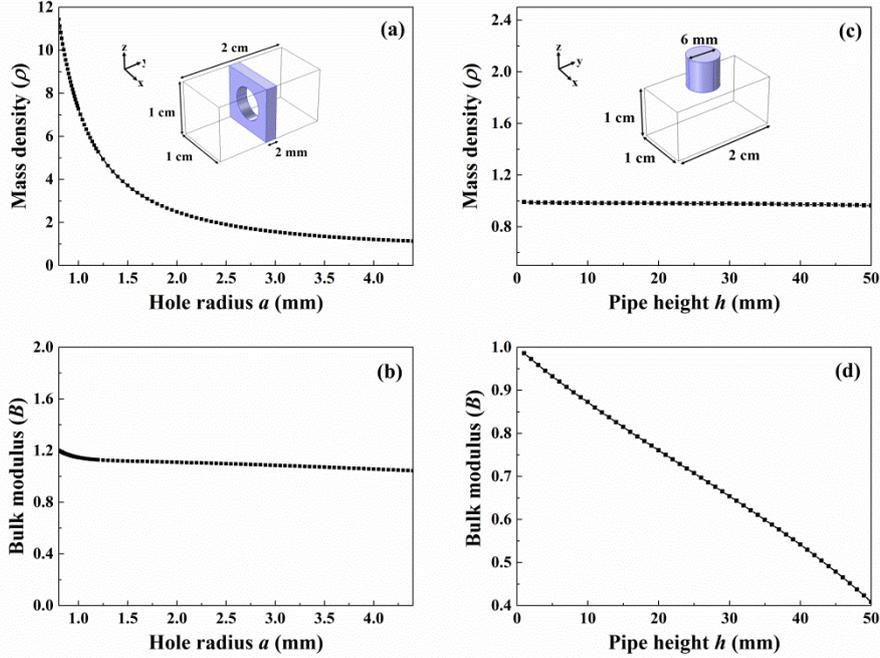

FIG. 2. Relationship curves for (a) the mass density and (b) the bulk modulus with the hole radius of the unit cell; the schematic of the unit cell with perforated plate is illustrated as the inset in (a). Relationship curves for (c) the mass density and (d) the bulk modulus with the pipe height of the unit cell; the schematic of the unit cell with side pipe is illustrated as the inset in (c).

In our previous research, the unit cell with perforated plate (inset of Fig. 2(a)) is used to modulate the mass density of the media by changing the radius of the hole.[29] From the work of Sam Hyeon Lee *et al.*, it is known that the unit cell with side pipe (inset of Fig. 2(c)) could be used to modulate the bulk modulus of the media by changing the height of the side pipe.[5] Then the effective parameters of these two unit cells are calculated using the well-developed retrieving method[30] to illustrate the characteristics of them. The geometry of each unit cell is a cuboid with length equal to 20 mm, width and height equal to 10mm. In the unit cell with perforated plate, the thickness of the perforated plate is 2 mm and the radius of the hole is $a$. In the unit cell with side pipe, the radius of the side pipe is 3mm and the height of the pipe is $h$. The sound waves are supposed to travel in y direction, namely circumferential direction of the acoustic bending structure. So the variations of refractive index and impedance of the unit cell in the orthogonal x direction have no influence on acoustic propagation. And only the acoustic parameters of y direction need to be modulated.

We study the acoustic parameters of the unit cell with perforated plate at first. The relationship between the mass density and the radius of the hole is shown in Fig. 2(a) and the relationship between the bulk modulus and the radius of the hole is shown in Fig. 2(b). It is clear that the mass density decreases from 11.4 to 1.1 when the radius of the hole varies from 0.8mm to 4.4 mm. Meanwhile, the bulk modulus remains almost constant around 1. Then we study the acoustic parameters of the unit cell with side pipe. The relationship between the mass density and the height of the pipe is shown in Fig. 2(c) and the relationship between the bulk modulus and the height of the pipe is shown in Fig. 2(d). It can be seen that the mass density of the unit cell with side pipe nearly keeps the same around 1 when the height of the pipe varies from 1mm to 50mm. However, the bulk modulus decreases from 0.99 to 0.41 when the height changes from 1mm to 50mm. These results verify that the unit cell with perforated plate can modulate the mass density and the unit cell with side pipe can modulate the bulk modulus separately. So we combine these two kinds of unit cells together to get a new complex unit cell shown in Fig. 3(c) to modify the mass density and bulk modulus simultaneously.

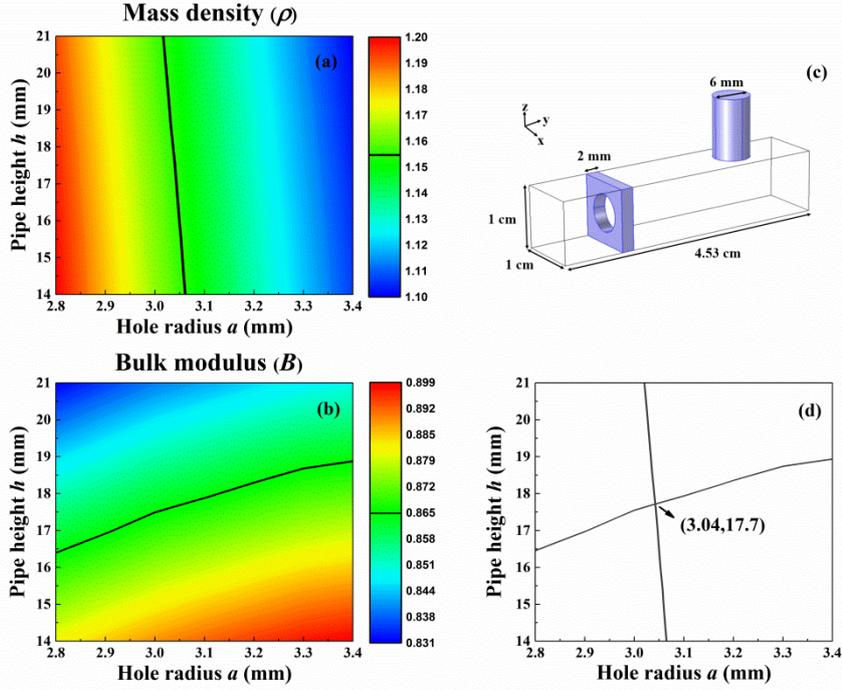

FIG. 3. Distribution maps of (a) the mass density and (b) the bulk modulus with different radii of the hole and different heights of the pipe. (c) Schematic of the complex unit cell. (d) Overlapping map of Figs. 3(a) and (b).

Although the perforated plate and the side pipe influence each other weakly, the mapping method to scan the radius of the hole and the height of the pipe at the same time is introduced for obtaining the exact geometry parameters of the complex unit cells. The process of calculating the geometry parameters of the unit cells whose radial position $r$ equals to 0.865m is given as an example. The distribution map of mass density with different radii of the hole and different heights of the pipe is shown in Fig. 3(a). The distribution map of bulk modulus is shown in Fig. 3(b). The x axis presents the variation of the radius of the hole and the y axis presents the variation of the height of the pipe. The black lines in the maps indicate the required values of mass density $\rho = 1.156$ and bulk modulus $B = 0.864$, which is obtained from Eq. (3). It's easy to get the exact geometry parameters by overlapping these two maps as shown in Fig. 3(d). The coordinate values of the intersection of the two black lines are precisely the required parameters. The rainbow color for other parameter values in Figs. 3(a) and (b) is set to transparent to make the overlapping map clear. From Fig. 3(d), it is observed that the values of the intersection is $a = 3.04$mm and $h = 17.7$mm. So the geometry parameters of the unit cells $r$=0.865m (i.e. the unit cells of 47$^{th}$ layer) are $a = 3.04$mm and $h = 17.7$mm.

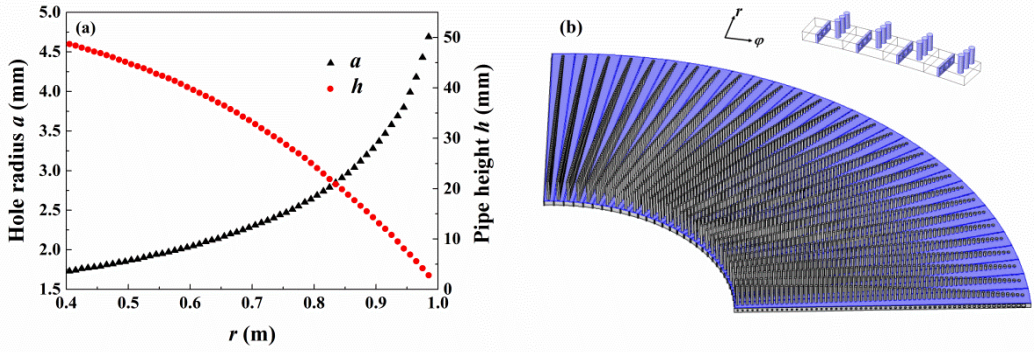

FIG. 4. (a) Distributions of the hole radius $a$ and the pipe height $h$ with the radial position $r$. (b) Schematic of the practical structure of the acoustic bend.

The different unit cells of other 58 layers along the radial direction are retrieved one by one using the same method mentioned above. The details of the geometry parameters of unit cells in each layer are shown in Fig. 4(a). The x axis stands for the radial positions; the left y axis stands for the radii of the perforated plates and the right y axis stands for the heights of the side pipes. From the center to the periphery along the radial direction, the radii of the holes increase from 1.73mm to 4.69mm indicated by red circle dots; the heights of the side pipes reduce from 48.7mm to 2.8mm indicated by black triangle dots. The whole structure of the IMAB configured with all the calculated unit cells is shown in Fig. 4(b). And the inset shows part of the structure composed of 12 unit cells. The maximal period $d$ of the outermost layer is 52mm which makes the unit cell at least 5 times smaller than the wavelength for frequencies under 1300Hz. Intuitively, this arrangement allows for smooth bending effect because a homogeneous medium can be assumed with criterion $d \leq \lambda/5$. The upper limit of the frequency can be easily extended to higher frequency by reducing the size of the unit cells, i.e. dividing the structure to more layers and more unit cells in each layer.

## Discussion

The performance of the IMAB was simulated by using the finite element method. An arc-shaped waveguide without acoustic bend was calculated firstly and presented in Fig. 5(a). The height of the waveguide is 10mm, and the width is 590mm. A plane wave was emitted from the left boundary and normally entered into the bend area. However, the sound mode was distorted when passing through the bend area. It is derived from undesired reflection on the hard wall. In Fig. 5(b), the IMAB was put in the bend area and the geometry parameters of the IMAB configured for the simulation were the same as the parameters given in Fig. 4(a). The incident surface and exit surface of the IMAB are indicated by the black dash lines. From the sound pressure field distribution in Fig. 5(b), it is observed that that the IMAB protects the mode in the waveguide effectively and guarantees the shape of the wave front when sound propagates through bend area.

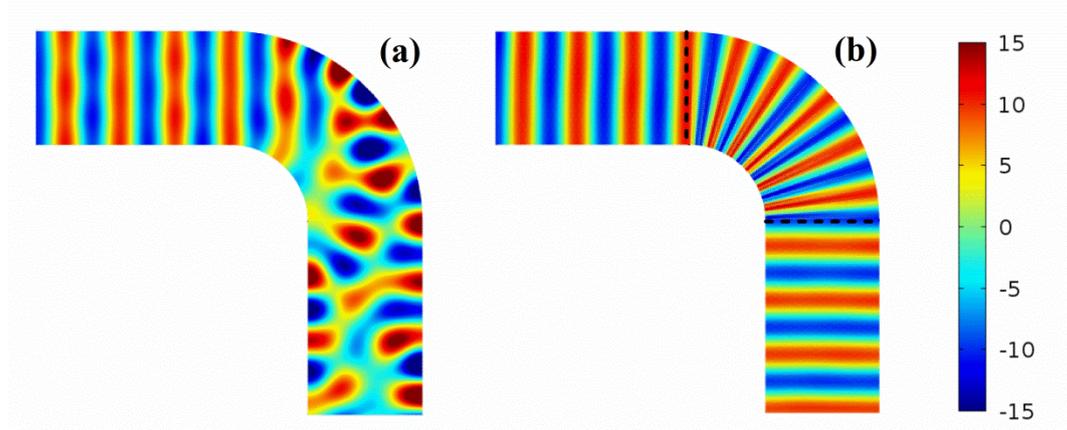

FIG. 5. Sound pressure field distribution at 1200Hz of (a) empty bending waveguide, (b) bending waveguide with the IMAB; the black dash lines stand for the incident surface and exit surface of the IMAB.

Then the energy transmission of the IMAB was calculated and presented as black square line in Fig. 6. It can be seen that the transmission keeps high at a wide frequency range which is derived from the impedance match. For comparison, the transmission curve for the ABPP was also calculated and presented as red circle line in Fig. 6. The transmission of the ABPP decreases because of the reflection at incident surface and exit surface for impedance mismatch. It is observed that the transmission of the IMAB structure was raised remarkably comparing to the ABPP. The transmission dip of the former is 0.93 while the transmission dip of the latter is 0.72, increased by 29%.

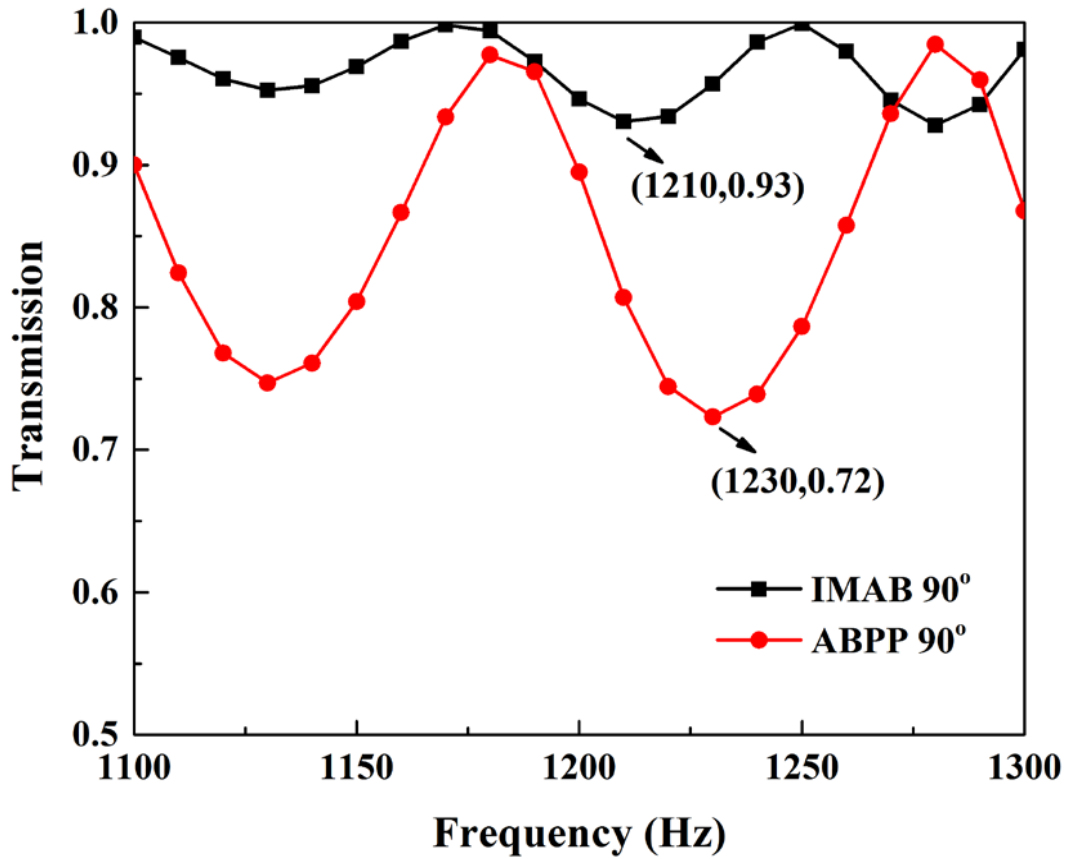

FIG. 6. Transmission curves of the IMAB and the ABPP.

The transmission curves fluctuate because of Fabry-Perot resonance. When sound waves propagate through an interlayer, the energy transmission could be expressed as follows:

$$t_I = \frac{4}{4\cos^2(k_2 D) + (R_{12} + R_{21})^2 \sin^2(k_2 D)},$$

$$R_{12} = \frac{R_2}{R_1} \quad R_{21} = \frac{R_1}{R_2}$$

(4)

$R_1$, $R_2$ are the impedance of background media and the interlayer media respectively; $D$ is the length of the interlayer. In this work, $R_1$, $R_2$ are the impedance of air and the acoustic bend; $D$ is the effective length of the acoustic bend in view of the arc shape of the structure. The transmission dip will appear when the length of the interlayer equals to odd times of the 1/4 sound wavelength and the transmission can be simplified as follows:

$$t_I = \frac{4}{(R_{12} + R_{21})^2}.$$

(5)

Substituting the value of the transmission dip into Eq. (5), the impedance ratio $R_{12}$ can be calculated easily. The impedance ratio for the ABPP is 1.80 while the impedance ratio for the IMAB is 1.31. So the impedance ratio improves by 27%.

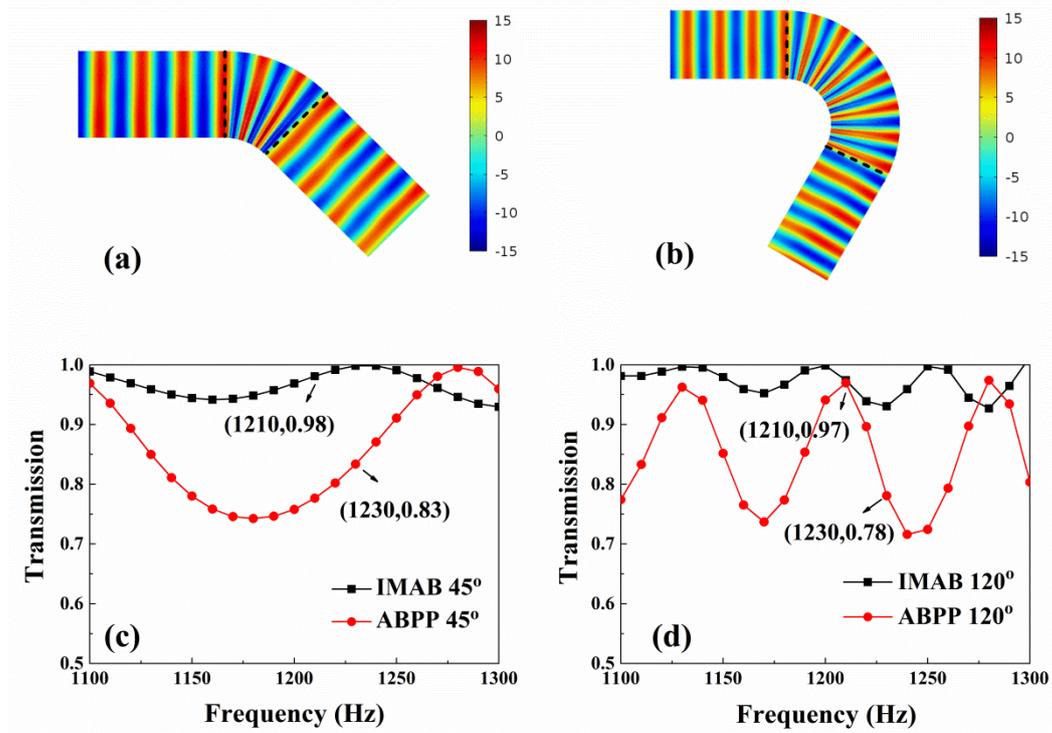

FIG. 7. Sound pressure field distribution at 1200Hz of the IMAB with angle of (a) 45° and (b) 120°. Transmission curves for acoustic bends with angle of (c) 45° and (d) 120°.

The IMABs with angle of 45° and 120° were also simulated in this work. The bending effect keeps as good as the IMAB with angle of 90° which can be observed from the sound pressure field distributions shown in Figs. 7(a) and (b). The mode of plane wave keeps perfectly after the sound passes through the bend area with different angles. The energy transmission curves for acoustic bends with angle of 45° and 120° are shown in Figs. 7(c) and (d). The transmission of the IMAB still keeps much higher than the transmission of the ABPP. For a single frequency, such as 1210Hz for the IMAB and 1230Hz for the ABPP, the transmissions for IMABs with different angles change little and keep higher than 0.9 while the transmissions for ABPPs with different angles change a lot for severe impedance mismatch. It indicates that the IMAB works effectively in a wide angle range.

## Conclusion

In conclusion, a metameterial based IMAB is designed and simulated. Complex unit cell with perforated plate and side pipe is utilized in the design to manipulate the mass density and the bulk modulus of media simultaneously. The well-developed retrieving method is used to obtain the effective mass density and bulk modulus of the unit cells and the mapping method is introduced to obtain the exact geometry parameters of the complex unit cells. The simulation results of the acoustic bends show that the IMAB protects the mode in the waveguide and guarantees the shape of the wave front effectively. And the great improvement of the impedance ratio of the IMAB keeps high energy transmission spectra in a wide frequency range and a wide angle range. This work once again shows the prospect of the metamaterial technology and the potential applications it would produce in the future.

## Acknowledgments

This work is supported by the Youth Innovation Promotion Association CAS (Grant No. 2017029) and the National Natural Science Foundation of China (Grant